# COMPARISON OF PREDICTIVE VALUES WITH PAIRED SAMPLES

*Martín Andrés, A.*[*] *and Femia Marzo, P.*

*Biostatistics. School of Medicine. University of Granada. 18071 Granada. Spain.*

## SUMMARY

Positive predictive value and negative predictive value are two widely used parameters to assess the clinical usefulness of a medical diagnostic test. When there are two diagnostic tests, it is recommendable to make a comparative assessment of the values of these two parameters after applying the two tests to the same subjects (paired samples). The objective is then to make individual or global inferences about the difference or the ratio of the predictive value of the two diagnostic tests. These inferences are usually based on complex and not very intuitive expressions, some of which have subsequently been reformulated. We define the two properties of symmetry which any inference method must verify - symmetry in diagnoses and symmetry in the tests -, we propose new inference methods, and we define them with simple expressions. All of the methods are compared with each other, selecting the optimal method: (a) to obtain a confidence interval for the difference or ratio; (b) to perform an individual homogeneity test of the two predictive values; and (c) to carry out a global homogeneity test of the two predictive values.

---

[*] Correspondence to: Bioestadística. Facultad de Medicina. Universidad de Granada. 18071 Granada. Spain. Email: amartina@ugr.es.

# 1. Introduction

A medical diagnostic test $A$ allows us to classify an individual as diseased (positive diagnosis, $A=+$) or as non-diseased (negative diagnosis, $A=-$). However, the reality, determined by a gold standard $S$, can be different: $S=+$ or $S=-$ whether or not the individual is really diseased, respectively. The evaluation of the quality of the $A$ test can be performed based on various parameters (sensitivity, specificity, diagnostic likelihood ratio, etc.), but its clinical usefulness must be assessed through the positive predictive values $P_A$ and negative $N_A$[1], i.e., the proportions of correct answers between positive or negative diagnoses, respectively. The usual notation for these two parameters is $PPV_A$ and $NPV_A$ respectively, but in what follows the previous notation ($P_A$ and $N_A$) will be used in order to shorten the formulae.

When another diagnostic test $B$ is available, with predictive values $P_B$ and $N_B$, the problem arises of the comparison of tests $A$ and $B$ through their respective predictive values. To this end, researchers often look at the difference parameters ($d=P_A-P_B$ and $\bar{d}=N_A-N_B$) or ratio ($R=P_A/P_B$ and $\bar{R}=N_A/N_B$) of the predictive values of both tests. Inferences about these parameters can be individual or global. In individual inferences, there may be three objectives: *(i)* to obtain a confidence interval (*CI*) for $d$, $\bar{d}$, $R$, or $\bar{R}$; *(ii)* perform an individual homogeneity test to check the hypothesis $H_d$: $d=0$ *vs.* $K_d$: $d\neq 0$ or $H_R$: $R=1$ *vs.* $K_R$: $R\neq 1$, which are equivalent hypotheses, and similarly for $\bar{d}$ or $\bar{R}$; and *(iii)* perform an individual non-inferiority test for hypotheses $H_{d(\delta)}$: $d=\delta$ *vs.* $K_{d(\delta)}$: $d>\delta$ (with $\delta<0$) or $H_{R(\rho)}$: $R=\rho$ *vs.* $K_{R(\rho)}$: $R>\rho$ (with $\rho<1$), and similarly for $\bar{d}$ or $\bar{R}$. In global inferences the objectives can be two: *(i)* obtain a confidence region (*CR*) for ($d$, $\bar{d}$) or ($R$, $\bar{R}$); and *(ii)* perform a two-tailed global homogeneity test for the null hypotheses $H_{d\bar{d}}$: $d=\bar{d}=0$ or $H_{R\bar{R}}$: $R=\bar{R}=1$, which are also equivalent hypotheses. These global tests are recommended as a previous step to the performance of individual tests[2]. All these inferences can be made based on two paired samples (cohort study) or based on two independent samples (case-control studies), depending on whether diagnostic tests *A* and *B* are applied in the same or different individuals, respectively. This article focuses on the case of two paired samples, which is most appropriate when comparing two alternative tests[3,4]. As a motivating example, the following refers to the classic example of Weiner *et al.*[5] whose data are found in Table 1(a). In this example –which is also alluded to by other authors[6,7,8]– the gold standard *S* for coronary artery disease is the result of coronary angiography in a multi-centre clinical trial, and the diagnostic tests are the result of exercise stress testing (test *B*) and clinical history of chest pain (test *A*).

Inferences about the parameters difference and ratio of the predictive values with paired samples were not well developed until 2006[3,9]. Until then, the traditional solution for individual inferences was the sophisticated and complex marginal regression method of Leisering *et al.*[6,10]. Less complicated, but just as effective as or even more effective than the previous ones, are the solutions of Wang *et al.*[7] for the cases of difference and ratio, Moskowitz and Pepe[9] for the case of ratio, and Kosinski[8] for the case of difference. This last article has the added advantage of providing more simplified and intuitive expressions of the variance of the different estimators, expressions that were too complex in the original format of Wang *et al.*[7] and Moskowitz and Pepe[9]. Regarding global inferences, Moskowitz and Pepe[9] and Roldán *et al.*[2] have addressed the parameters ratio or difference, respectively, but in both cases the expressions of covariances are also complex and not very intuitive; this article aims to correct that problem. Additionally, Tsou[11] has introduced a "robust score test statistic" that, as it has fewer nuisance parameters, produces identical results as those of the authors cited above.

Most of the above inferences about the parameters $d$, $\bar{d}$, $R$, and $\bar{R}$ are based on the application of the delta method to the statistics $\hat{d}$, $\hat{\bar{d}}$, $\hat{R}$, and $\hat{\bar{R}}$ -the maximum likelihood estimators of the parameters $d$, $\bar{d}$, $R$, and $\bar{R}$, respectively-, which may not work well in certain circumstances (see Martín Andrés and Álvarez Hernández[12], for the case of the ratio of two independent proportions). In all cases, except for the exception already mentioned by Kosinski[8], the expressions obtained are usually complex and not very intuitive. This is due to the expression of the matrix obtained by the delta method and mainly because certain sums of probabilities maintained in the demonstrations could be replaced by functions of the corresponding predictive values. Our current objectives are four: (1) to propose corrections to known inference methods; (2) to propose new methods; (3) to express in a simple format all known and new expressions that are obtained; and (4) to compare by simulation some of the individual or global inferences of the old and new methods (*CI*, individual homogeneity test, and global homogeneity test). To achieve several of these objectives, the symmetry properties presented in the next section will be helpful.

**2. Model, parameters of interest and properties of symmetry**

Tables 1 (a) and (b) indicate the notation to be used for the frequencies and probabilities of a paired study design, respectively. Table 1 (a) also includes the numerical results of the classic example of Weiner *et al.*[5] referred to in Section 1. Data were obtained by classifying *n*=871 individuals as in Table 1(a); therefore $n = \sum_{i=1}^{8} x_i = \sum_{j=1}^{4} n_j = 871$, with $n_j = x_j + x_{j+4}$ (*j*=1 to 4), and ($x_1, \ldots, x_8$) refers to a multinomial distribution M{*n*; $p_1, \ldots, p_8$} with $\sum_{i=1}^{8} p_i = \sum_{j=1}^{4} t_j = 1$ and $t_j = p_j + p_{j+4}$ (*j*=1 to 4). The four predictive values are given by the expressions

$$P_A = \frac{p_1 + p_2}{t_1 + t_2} = \frac{p_A}{t_A}, \; P_B = \frac{p_1 + p_3}{t_1 + t_3} = \frac{p_B}{t_B}, \; N_A = \frac{p_7 + p_8}{t_3 + t_4} = \frac{\overline{p_A}}{\overline{t_A}}, \text{ and } N_B = \frac{p_6 + p_8}{t_2 + t_4} = \frac{\overline{p_B}}{\overline{t_B}},$$

and their estimators are

$$\hat{P}_A = \frac{x_1 + x_2}{n_1 + n_2} = \frac{x_A}{n_A}, \; \hat{P}_B = \frac{x_1 + x_3}{n_1 + n_3} = \frac{x_B}{n_B}, \; \hat{N}_A = \frac{x_7 + x_8}{n_3 + n_4} = \frac{\overline{x_A}}{\overline{n_A}}, \text{ and } \hat{N}_B = \frac{x_6 + x_8}{n_2 + n_4} = \frac{\overline{x_B}}{\overline{n_B}},$$

respectively. For the example in Table 1(a) we obtain $\hat{P}_A = 0.8935$, $\hat{P}_B = 0.8807$, $\hat{N}_A = 0.7849$,

and $\hat{N}_B = 0.6478$. Note that each of these estimators can be considered a binomial proportion with denominators $n_A$, $n_B$, $\bar{n}_A$, and $\bar{n}_B$ respectively, i.e. the total number of positive or negative diagnoses of tests $A$ and $B$, respectively. With this notation, $p_A = p_1 + p_2 = t_A P_A$, $x_A = x_1 + x_2 = n_A \hat{P}_A$, $\hat{t}_A = n_A/n$, etc., which will be used frequently in the demonstrations of the Appendices in order to simplify the final expressions. The point estimators of the four parameters of interest are $\hat{d} = \hat{P}_A - \hat{P}_B = 0.0128$, $\hat{\bar{d}} = \hat{N}_A - \hat{N}_B = 0.1370$, $\hat{R} = \hat{P}_A / \hat{P}_B = 1.015$ and $\hat{\bar{R}} = \hat{N}_A / \hat{N}_B = 1.212$.

Many authors often offer different demonstrations and formulae for statistics involving $P_A$ and $P_B$ parameters on the one hand, and $N_A$ and $N_B$ on the other. In fact, the latter can be deduced from the former -thus avoiding unnecessary duplication- if the following property of *symmetry in diagnoses* is applied. If Table 1(a) permutes the signs + and −, i.e. if the new + and − refer to "non-diseased" and "diseased" respectively, then the new $P_X$ values (with $X=A$ or $B$) are the old $N_X$ values, and the old data $(x_1, \ldots, x_8)$, $(n_1, n_2, n_3, n_4)$, $(x_A, x_B, \bar{x}_A, \bar{x}_B)$, and $(n_A, n_B, \bar{n}_A, \bar{n}_B)$ are rearranged as $(x_8, \ldots, x_1)$, $(n_4, n_3, n_2, n_1)$, $(\bar{x}_A, \bar{x}_B, x_A, x_B)$, and $(\bar{n}_A, \bar{n}_B, n_A, n_B)$, respectively. We can do the same with the parameters in Table 1(b). A consequence of this property is that any proof or formula that alludes to estimators $\hat{P}_X$, is also valid for estimators $\hat{N}_X$ if it changes the values

$$(x_i, n_j, x_X, \bar{x}_X, n_X, \bar{n}_X, \hat{P}_X, \hat{N}_X)$$

for the values

$$(x_{9-i}, n_{5-j}, \bar{x}_X, x_X, \bar{n}_X, n_X, \hat{N}_X, \hat{P}_X),$$

respectively, where $i=1$ to 8, $j=1$ to 4, and $X=A$ or $B$. Likewise for the $P_X$ and $N_X$ parameters: $(p_i, t_j, p_X, \bar{p}_X, t_X, \bar{t}_X, P_X, N_X)$ are replaced by $(p_{9-i}, t_{5-j}, \bar{p}_X, p_X, \bar{t}_X, t_X, N_X, P_X)$, respectively.

As an example, it is immediate to verify that with these changes the formula of $\hat{P}_A$ becomes the formula of $\hat{N}_A$. That is why in everything that follows, frequently only the expressions related to $P_X$ will be provided, omitting those related to $N_X$.

For the inferences to be consistent, it is also necessary that the formulae obtained verify the following property of *symmetry in the tests*. If the *A* and *B* tests are permuted in Table 1, the inferences about the new values $d'$, $\bar{d}'$, $R'$, or $\bar{R}'$ must be compatible with the inferences about the old values $d$, $\bar{d}$, $R$, or $\bar{R}$. For example, if the *CI* for $d$ is $(d_L, d_U)$, the *CI* for $d'$ must be $(-d_U, -d_L)$; similarly, if the *CI* for $R$ is $(R_L, R_U)$, the *CI* for $R'$ must be $(1/R_U, 1/R_L)$. Jamart[13] outlined this aspect in the case of the homogeneity test.

Finally, it is noteworthy that Bennett[14] suggested an individual homogeneity test, but it does not verify the symmetry properties in the test[7,13] or in diagnoses, as it can be proved easily. This comes from several errors in the establishment of the test statistics for $H_d$: $d=0$ and $H_{\bar{d}}$: $\bar{d}=0$. Wu[15] has proved that if the errors in the Bennett statistic are corrected, we can obtain a statistic that is asymptotically equivalent to the Wang *et al.*[7] statistic. In Appendix B the Bennett statistic is analyzed from several perspectives.

### 3. Inferences about differences of predictive values

Leisenring *et al.*[6] define the marginal regression model that allows to make different inferences, providing particularly appropriate statistics for the individual homogeneity tests. The method is complex, but the formulae were simplified by Roldán *et al.*[2] and especially by Kosinski[8]. The latter proposes an improved versión of the test statistic of Leisering *et al.* for hypothesis $H_d$: $P_A=P_B(=P)$; the resulting statistic -weighted generalized score statistic- is called $z^2_{d(p)}$ in Section 3.

Applying the Multivariate Central Limit Theorem (MCLT) together with the delta method to the vector $(\hat{d}, \hat{\bar{d}})$, in Appendix A it is shows that asymptotically $(\hat{d}-d, \hat{\bar{d}}-\bar{d})$ $\sim N\{(0,0); \hat{\Sigma}_d\}$, where $\hat{\Sigma}_d$ is the variance-covariance matrix given by expressions

$$\hat{\sigma}_d^2 = \frac{\hat{P}_A(1-\hat{P}_A)}{n_A} + \frac{\hat{P}_B(1-\hat{P}_B)}{n_B} - 2\frac{(1-\hat{P}_A)(1-\hat{P}_B)x_1 + \hat{P}_A\hat{P}_B x_5}{n_A n_B}, \qquad (1)$$

$$\hat{\sigma}_{\bar{d}}^2 = \frac{\hat{N}_A(1-\hat{N}_A)}{\bar{n}_A} + \frac{\hat{N}_B(1-\hat{N}_B)}{\bar{n}_B} - 2\frac{(1-\hat{N}_A)(1-\hat{N}_B)x_8 + \hat{N}_A\hat{N}_B x_4}{\bar{n}_A \bar{n}_B}, \qquad (2)$$

$$\hat{\sigma}_{d\bar{d}} = \frac{(1-\hat{P}_A)\hat{N}_B x_2 + \hat{P}_A(1-\hat{N}_B)x_6}{n_A \bar{n}_B} + \frac{(1-\hat{P}_B)\hat{N}_A x_3 + \hat{P}_B(1-\hat{N}_A)x_7}{\bar{n}_A n_B}. \qquad (3)$$

Note that expression (2) can be obtained from expression (1) by applying to the latter the property of symmetry in diagnoses, while expression (3) does not change when applying the mentioned property. Expressions (1) and (2) are not the original and complex expressions of Wang et al.[7], owing its current format to Kosinski[8]; expression (3) is new, to the best of our knowledge. Note also that the first two addends of expression (1) correspond to the classical Wald estimator of the variance of the difference of two independent proportions; the third addend is $-2\widehat{\text{Cov}}\{\hat{P}_A, \hat{P}_B\}$. From the former, we can obtain several inferences. When the goal is to make individual inferences on $d$, then the Wang et al.[7] expression is obtained: the CI for $d$ and the test statistics for the $H_{d(\delta)}$ and $H_d$ hypothesis are, respectively

$$CI_d: d \in \hat{d} \pm z_\alpha \hat{\sigma}_d, \quad z_{d(\delta)}^2 = (\hat{d}-\delta)^2 / \hat{\sigma}_d^2, \text{ and } z_d^2 = \hat{d}^2 / \hat{\sigma}_d^2, \qquad (4)$$

where $z_\alpha$ is the $(1-\alpha/2)$-percentile of the typical normal distribution, $z_{d(\delta)}$ is compared with $+z_{2\alpha}$, and $z_d^2$ is compared with $z_\alpha^2$. The $CI_d$ is achieved by solving $\delta$ in the equation $z_{d(\delta)}^2 = z_\alpha^2$. When the goal is to make global inferences on the differences, we obtain the Roldán et al.[2] method: the CR for $(d, \bar{d})$ and the test statistic for $H_{d\bar{d}}$ are given by

$$(\hat{d}-d, \hat{\bar{d}}-\bar{d})\hat{\Sigma}_d^{-1}(\hat{d}-d, \hat{\bar{d}}-\bar{d})' \leq \chi_{2,\alpha}^2 \text{ and } \chi_d^2 = (\hat{d},\hat{\bar{d}})\hat{\Sigma}_d^{-1}(\hat{d},\hat{\bar{d}})', \qquad (5)$$

respectively, where $\chi_{2,\alpha}^2$ the $(1-\alpha)$-percentile of the chi-square distribution with 2 df and $\chi_d^2$ is compared with $\chi_{2,\alpha}^2$. Note that the last expression of (5) is simpler than the Roldán et al.[2]

expression, and is equivalent to this if the errors in the latter are corrected.

In the next two sections, we will avoid repeating many of the aspects reviewed in the section- verification of symmetry, creation of the non-inferiority test, theoretical quantities, etc- restricting ourselves to only indicating the novelties in each case.

## 4. Inferences about ratios of predictive values

In Appendix A it is also proved that asymptotically ($\ln \hat{R} - \ln R$, $\ln \hat{\bar{R}} - \ln \bar{R}$) ~ $N\{(0,0); \hat{\Sigma}_R\}$, where $\hat{\Sigma}_R$ is the matrix given by

$$\hat{\sigma}_R^2 = \frac{1-\hat{P}_A}{n_A \hat{P}_A} + \frac{1-\hat{P}_B}{n_B \hat{P}_B} - \frac{2}{n_A n_B} \left\{ \frac{(1-\hat{P}_A)(1-\hat{P}_B)}{\hat{P}_A \hat{P}_B} x_1 + x_5 \right\}, \quad (6)$$

$$\hat{\sigma}_{\bar{R}}^2 = \frac{1-\hat{N}_A}{\bar{n}_A \hat{N}_A} + \frac{1-\hat{N}_B}{\bar{n}_B \hat{N}_B} - \frac{2}{\bar{n}_A \bar{n}_B} \left\{ \frac{(1-\hat{N}_A)(1-\hat{N}_B)}{\hat{N}_A \hat{N}_B} x_8 + x_4 \right\}, \quad (7)$$

$$\hat{\sigma}_{R\bar{R}} = \frac{1}{n_A \bar{n}_B} \left\{ \frac{1-\hat{P}_A}{\hat{P}_A} x_2 + \frac{1-\hat{N}_B}{\hat{N}_B} x_6 \right\} + \frac{1}{\bar{n}_A n_B} \left\{ \frac{1-\hat{P}_B}{\hat{P}_B} x_3 + \frac{1-\hat{N}_A}{\hat{N}_A} x_7 \right\}, \quad (8)$$

Expressions (6) and (7) are not the original and complex expressions of Wang et al.[7] and Moskowitz and Pepe[9], and owe their current format to Kosinski[8]. Expression (8) is a new format of the complex expression of Moskowitz and Pepe[9]. Note also that the first two addends of expression (6) correspond to the classical estimator of Wald of the variance of the logarithm of the ratio of two independent proportions. The third addend is $-2\widehat{\text{Cov}}\{\log \hat{P}_A, \log \hat{P}_B\}$. From the former we obtain the following inferences. The *CI* for *R* and the test statistics for $H_{R(\rho)}$ and $H_R$ are the statistics of Wang et al.[7]

$$CI_{LR}: R \in \hat{R} \times \exp\{\pm z_\alpha \hat{\sigma}_R\}, \quad z_{LR(\rho)}^2 = (\log \hat{R} - \log \rho)^2 / \hat{\sigma}_R^2 \quad \text{and} \quad z_{LR}^2 = \ln^2 \hat{R} / \hat{\sigma}_R^2, \quad (9)$$

while the *CR* for $(R, \bar{R})$ and the test statistic for $H_{R\bar{R}}$[9] will be , respectively

$$\left(\log \frac{\hat{R}}{R}, \log \frac{\hat{\bar{R}}}{\bar{R}}\right) \hat{\Sigma}_R^{-1} \left(\log \frac{\hat{R}}{R}, \log \frac{\hat{\bar{R}}}{\bar{R}}\right)' \leq \chi_{2,\alpha}^2 \quad \text{and} \quad \chi_{LR}^2 = \left(\log \hat{R}, \log \hat{\bar{R}}\right) \hat{\Sigma}_R^{-1} \left(\log \hat{R}, \log \hat{\bar{R}}\right)'. \quad (10)$$

At the end of Appendix A it can be seen the importance of the new expressions of the variance for the establishment the sample size based on the formulae of Wang et al.[7] and Moskowitz and Pepe[9].

The former inferences are based on the natural logarithm of estimators $\hat{R}$ or $\hat{\bar{R}}$. To obtain inferences based directly on $\hat{R}$ or $\hat{\bar{R}}$, in Appendix A it is shown how to modify the inferences of the former paragraph. So, we can get the following new expressions

$$CI_R : R \in \hat{R} \times \left[ Y \pm \sqrt{Y^2 - 1} \right] \text{ with } Y = 1 + \frac{z_\alpha^2 \hat{\sigma}_R^2}{2}, \; z_{R(\rho)}^2 = \frac{(\hat{R} - \rho)^2}{\rho \hat{R} \hat{\sigma}_R^2}, \; z_R^2 = \frac{(\hat{R} - 1)^2}{\hat{R} \hat{\sigma}_R^2}, \quad (11)$$

$$\left( \frac{\hat{R} - R}{\sqrt{\hat{R}R}}, \frac{\hat{\bar{R}} - \bar{R}}{\sqrt{\hat{\bar{R}}\bar{R}}} \right) \hat{\Sigma}_R^{-1} \begin{pmatrix} \frac{\hat{R} - R}{\sqrt{\hat{R}R}} \\ \frac{\hat{\bar{R}} - \bar{R}}{\sqrt{\hat{\bar{R}}\bar{R}}} \end{pmatrix} \leq \chi_{2,\alpha}^2 \text{ and } \chi_R^2 = \left( \frac{\hat{R} - 1}{\sqrt{\hat{R}}}, \frac{\hat{\bar{R}} - 1}{\sqrt{\hat{\bar{R}}}} \right) \hat{\Sigma}_R^{-1} \begin{pmatrix} \frac{\hat{R} - 1}{\sqrt{\hat{R}}} \\ \frac{\hat{\bar{R}} - 1}{\sqrt{\hat{\bar{R}}}} \end{pmatrix}. \quad (12)$$

## 5. Modifications of the previous inference procedures

All of the homogeneity tests described in Sections 3 and 4 are based on the Wald method, which means that the variances are estimated by substituting $p_i$ parameters for their unrestricted maximum likelihood estimators $\hat{p}_i = x_i/n$. It may be more convenient to estimate predictive values under the null hypothesis of their equality. As in this case it occurs that $P_A = P_B = P$ (or $N_A = N_B = N$), then $P_A$ and $P_B$ (or $N_A$ and $N_B$) can be estimated by a single value $\hat{P}$ (or $\hat{N}$) given by the weighted estimators

$$\hat{P} = \left( n_A \hat{P}_A + n_B \hat{P}_B \right) / (n_A + n_B) \text{ and } \hat{N} = \left( \bar{n}_A \hat{N}_A + \bar{n}_B \hat{N}_B \right) / (\bar{n}_A + \bar{n}_B). \quad (13)$$

By making $\hat{P}_A = \hat{P}_B = \hat{P}$ in statistics $z_d^2$ and $z_{LR}^2$ of expressions (4) and (9) respectively, we obtain statistics

$$z_{d(p)}^2 = \frac{\hat{d}^2}{\hat{\sigma}_{d(p)}^2} \text{ with } \hat{\sigma}_{d(p)}^2 = \hat{P}(1 - \hat{P}) \times \left\{ \frac{1}{n_A} + \frac{1}{n_B} \right\} - 2 \frac{(1 - \hat{P})^2 x_1 + \hat{P}^2 x_5}{n_A n_B}, \quad (14)$$

$$z_{LR(p)}^2 = \frac{\ln^2 \hat{R}}{\hat{\sigma}_{R(p)}^2} \quad \text{with} \quad \hat{\sigma}_{R(p)}^2 = \frac{1-\hat{P}}{\hat{P}}\left\{\frac{1}{n_A}+\frac{1}{n_B}\right\} - \frac{2}{n_A n_B}\left\{\left(\frac{1-\hat{P}}{\hat{P}}\right)^2 x_1 + x_5\right\} \quad (15)$$

When both samples are independent, statistic $z_{d(p)}^2$ [8] has the advantage of becoming the classical statistic for the comparison between two independent proportions. This is the cause of having used subindex "d(p)": it is an statistic based on the "*difference (with pooled estimator)*". Doing the same with all of the statistics of the homogeneity test in Sections 3 and 4 we can obtain statistics $z_{LR(p)}^2$ and $z_{R(p)}^2$ for the individual tests, and $\chi_{d(p)}^2$, $\chi_{LR(p)}^2$, and $\chi_{R(p)}^2$ for the global test.

Finally, when using the logarithm transformation -as happens with the $CI_{LR}$ of expression (9) for case *R*- it is usual to use the expressions after increasing the data by 0.5[16]. As this increase by 0.5 has also proved to be effective in the cases of difference[17] and the ratio[12] of two independent proportions, our proposal is to apply this increase to all of the methods in Sections 3 and 4. Therefore, methods *d*, *LR* and *R* will lead to the new adjusted Wald *d(a)*, *LR(a)* and *R(a)* methods, respectively. For example, statistic $z_{LR}^2$ applied to the increased data in 0.5 will lead to statistic $z_{LR(a)}^2$.

## 6. Comparison of inferential methods by simulation.

### 6.1. General considerations

The goal of this section is to compare by simulation the individual and global homogeneity tests, as well as the *CI* proposed previously. In the following $\pi = p_1+p_2+p_3+p_4$ is the prevalence, and $O^+=p_1p_4/p_2p_3$ and $O^-=p_5p_8/p_6p_7$ are the degrees of association between the diagnoses of *A* and *B* in diseased and non-diseased individuals, respectively. Partially following Kosinski[8], every simulation is based on the incoming two steps: (1) we set values of $\{P_A, P_B, N_A, N_B, \pi, O^+, O^-\}$ and, from those values, we establish the parameters $\{p_1, p_2, ..., p_8\}$ through Table III of Kosinski[8]; and (2) we set the required size of the *n* sample and we extract *N* tables randomly of the multinomial M$\{n | p_1, p_2, ..., p_8\}$, tables that will be a basis for the evaluations. The values of $P_A$, $P_B$, $N_A$ and $N_B$ will be the ones specified in each case, $\pi$=0.35 or 0.65, $O^+$ and $O^-$=2 or 5, $n$=100, 200 or 300 and $N$=$10^7$, which guarantees the precision of at least the first decimal of the percentage that will be estimated. Additionally when we obtain some value for $x_i$=0, this is substituted by $x_i$=0.05 (for the same reason as Wang et al.[7]).

The following comparisons are restricted to the $d$ and $R$ cases, not taking into account cases $\bar{d}$ and $\bar{R}$. The reason once again is the property of symmetry in diagnoses, which is now complemented with $\{\pi, O^+, O^-\}$ being substituted by $\{1-\pi, O^-, O^+\}$, respectively. Suppose that we want to evaluate the CI for $\bar{d}$ based on the setting $\{P_A$=0.8, $P_B$=0.8, $N_A$=0.8, $N_B$=0.7, $\pi$=0.35, $O^+$=5, $O^-$=2$\}$. Because of the property of symmetry in diagnoses that evaluation is equivalent to evaluating the CI for $d$ based on the setting $\{P_A$=0.8, $P_B$=0.7, $N_A$=0.8, $N_B$=0.8, $\pi$=0.65, $O^+$=2, $O^-$=5$\}$. If simulations are performed with these two symmetrical configurations, the conclusion will be that it is sufficient to assess the CI for parameter $d$. Additionally, the assessments do not take into account the pair ($P_A$=0.7, $P_B$=0.8), since due to the property of symmetry in the tests, their results are the same as in the pair ($P_A$=0.8, $P_B$=0.7) which is taken into account. A similar thing occurs with parameter $R$.

*6.2. Comparison of the confidence intervals*

Until now, different authors have only been concerned with comparing the different statistics of individual homogeneity tests that currently exist, but they have not evaluated the CIs for the parameters involved. Nevertheless, we believe that this latter question has greater interest, since what is relevant to choose test $A$ or test $B$ is the magnitude of the parameters $d$, $R$, $\bar{d}$, or $\bar{R}$. The current objective is to compare the two CIs for $d$ proposed in the previous sections -those which have the subindex $d$ and $d(a)$- and the four ones proposed for $R$ -those which have the subindex $LR$, $LR(a)$, $R$, and $R(a)$-. For this purpose, for the values $\{P_A, P_B, N_A, N_B, \pi, O^+, O^-, n\}$ of each line of Table 2 we generate $N$=$10^7$ tables. In each table $h$ ($h$=1, ...,

*N*) the *CI* required ($d_{hL}$; $d_{hU}$) is obtained for a confidence of 95%, $I_h$=1 (or 0) is made if the *CI* does (or does not) contain the value $d=P_A-P_B$ and the width $W_h=d_{hU}-d_{hL}$ del *CI* is calculated. Finally, we note down the values of $C=(\Sigma I_h/N)\times 100\%$ and $W=\Sigma W_h/N$: the empirical coverage $1-\alpha^*$ and the average width of the *CI*, respectively. The procedure is done in a similar way for the case of *R*. Table 2 shows the results of the six *CI*s proposed.

In the case of the parameter "difference of predictive values", Table 2 shows that method *d* has coverages (*C* values) between 90.0% and 94.9%, and they are generally lower than the nominal value of 95% in small prevalences. Nonetheless, method *d(a)*, which is usually somewhat conservative, has a minimum coverage of 94.7% and always has more coverage than method *d*. However, this greater coverage of method *d(a)* is not obtained at the cost of a greater value of *W*. It can be observed that the *W* values of method *d(a)* are frequently lower than or equal to those of method *d* and that, when they are larger, there is only a very small difference. The consequence is that in order to obtain *CI* one must use method *d(a)* and not method *d*.

In the case of the parameter "ratio of predictive values", Table 2 shows that in general the pairs of methods {*LR*, *R*} and {*LR(a)*, *R(a)*} are very similar. The values *C* and *W* of the first method of each pair are always greater than or equal to those of the second method, but the differences, when they exist, are usually small. All of the methods have a good average coverage, but methods *LR* and *R* can give *C* values as small as 92.8% or 92.7%, respectively, in small prevalences and small sample sizes. On the contrary, the methods *LR(a)* and *R(a)* have *C* values ≥94.7% and, although the C values of *LR(a)* are sometimes slightly higher than those of *R(a)*, this is not an advantage since this happens when *C*≥95%. The global conclusion is that *R(a)* is slightly better than *LR(a)* -although both methods are acceptable-, but methods *LR* and *R* must be ruled out. The two methods selected are slightly liberal (94.7%≤*C*<95%) on a few occasions in which the prevalence is high.

It should be noted that the previous analysis and conclusions have been made based on the average width. Our data indicate that there is no substantial variation through reasoning in the same way, but based on the median width.

*6.3. Comparison of the individual homogeneity tests*

The objective now is to assess the nine tests that have been defined to contrast $H$: $P_A=P_B$ ($\equiv H_d \equiv H_R$) vs. $K$: $P_A \neq P_B$ ($\equiv K_d \equiv K_R$): the nine tests are denoted as $d$, $d(a)$, $d(p)$, $LR$, $LR(a)$, $LR(p)$, $R$, $R(a)$, and $R(p)$. For this purpose, for the values $\{P_A, P_B, N_A, N_B, \pi, O^+, O^-, n\}$ of each line in Table 3 we have generated $N=10^7$ tables. In each table $h$ ($h=1, …, N$) the test required is carried out to a nominal error of $\alpha=5\%$, $I_h=1$ (or 0) is considered if the test is significant (or not significant), and the empirical size of the method is calculated: $\alpha^*=(\Sigma I_h/N) \times 100\%$. Table 3 shows the results for the nine tests proposed. As the *CI*s for $d$ or $R$ in the previous section are obtained through inversion of the two-tailed test for the null hypotheses $H_{d(\delta)}$ or $H_{R(\rho)}$, then the empirical size $\alpha^*$ of a homogeneity test is the value of "1−empirical coverage" of the *CI* from which it comes when $P_A=P_B$, i.e. when $d=0$ or $R=1$. This is why the values in Table 3 are equal to one hundred minus values of $C$ in Table 2, in the lines with $P_A=P_B$ and for the inference methods that figure in both tables. It should be noted that these simulations are wider than the traditional ones of Wang *et al.*[7] and Kosinski[8], since these authors carry them out based on settings in which $P_A=P_B$ and $N_A=N_B$ simultaneously, what should not happen in general. Something similar occurs with the simulations of Table 4, which are cited below.

Partially following the criterion of Kosinski[8], in the first phase it is advisable to select the methods that preserve the nominal error of 5% ($\alpha^* \leq 5\%$) or that do not exceed it by too much. It can be seen in Table 3 that: (1) the methods without "*a*" or "*p*" ($d$, $LR$ and $R$) must be ruled out, since in these $\alpha^*$ can reach 10%, 7.2% and 7.2% respectively; (2) two of the methods with "*p*" -$LR(p)$ and $R(p)$- can be accepted, since in these $\alpha^*$ is not greater than the values of 5.2% and 5.3% respectively, and only exceeds the nominal error of 5% on one or two occasions; and (3) the four remaining methods -*d(a)*, *LR(a)*, *R(a)* and *d(p)*- are clearly acceptable, since they verify that $\alpha^* \leq 5\%$. These results are in line with the conclusion reached by Kosinski[8] that method *d(p)* is better than methods *d* and *LR*. Wang *et al.*[7] concluded that method *d* is better than *LR*, a statement that does not make much sense under the current data, since *d* (or *LR*) has an empirical size between 5.1%-10% (or 4.2%-7.2%). In fact, the two tests perform badly.

Table 4 summarizes the empirical power $\Theta$ for the settings and methods that are indicated. Methods *d*, *LR* and *R* are not included in the table since their power would be rather inflated because of the excessive value of $\alpha^*$. The simulation procedure is the same as in the first paragraph of this section, but now $\Theta = (\Sigma I_h/N) \times 100\%$. It can be observed that:

(1) Among the methods that respect the error $\alpha=5\%$ -*d(a)*, *d(p)*, *LR(a)*, and *R(a)*- the best one globally is method *d(p)*, followed by method *d(a)*. In particular, *d(p)* is bettered by *R(a)* when $P_A - P_B = N_B - N_A$ and the prevalence is low.

(2) Among the methods that do not respect the error $\alpha=5\%$ by a very small margin -*LR(p)* and *R(p)*- *R(p)* should be selected as it always has either equal or greater power than *LR(p)*.

(3) The comparison of the two methods selected -*d(p)* and *R(p)*- indicates that *d(p)* is the best, and is only bettered by *R(p)* when $P_A - P_B = N_A - N_B$ and the prevalence is low. Nevertheless, *R(p)* competes well with *d(a)*, the second method selected in the first step.

The final conclusion is that *d(p)* is the best individual homogeneity test, followed by method *R(p)*, which is more powerful than the former only when $(P_A - P_B) = (N_A - N_B)$ and the prevalence is low. Moreover, if we wish the results of the homogeneity test and of the *CI* to be compatible, method *d(a)* is a good procedure for the case of the difference in risks. Nevertheless, method *R(a)* is not a good one in the case of the risk ratio, since generally it has less power than method *R(p)*.

*6.4. Comparison of the global homogeneity test*

To check that $H$: $(P_A=P_B)\cap(N_A=N_B)$ vs. $K$: "one of the two equalities is not true" nine statistics have been proposed: the three which are indicated in expressions (5), (10) and (12) - $\chi^2_d$, $\chi^2_{LR}$, and $\chi^2_R$ -, all three of them are "pooled" ones which are obtained from the aforementioned expressions using the estimators of expression (13) - $\chi^2_{d(p)}$, $\chi^2_{LR(p)}$, and $\chi^2_{R(p)}$ -, and all three of them are "adjusted" ones which are obtained by increasing the frequencies $x_i$ en 0.5 - $\chi^2_{d(a)}$, $\chi^2_{LR(a)}$, and $\chi^2_{R(a)}$ -. To the best of our knowledge, until now it has not been compared the performance of any pair of these statistics, not even that of the two classic statistics $\chi^2_d$ and $\chi^2_{LR}$. The simulation procedure is the same as in the previous section, but for the settings of Tables 5 (empirical size) and 6 (empirical power), although the statistics are compared with $\chi^2_{2,\ \alpha=5\%}$.

The data from Table 5 indicate that the classic method $d$ is always liberal, since its empirical size lies between 5.1% and 6.7%, and therefore it must be ruled out. All of the other methods have an empirical size that is never greater than the nominal error of 5%, and therefore they must be accepted. The data from Table 6 indicate that method $R$ always has a power equal to or greater than that of the other methods, and this is why it should be the one we select. Nevertheless, method $LR$ is only slightly worse. The differences between methods are more relevant the smaller $n$ is, but they are less relevant when $\{P_A<P_B,\ N_A>N_B\}$ or $\{P_A>P_B,\ N_A<N_B\}$.

**7. Examples**

Table 7 contains some inferences based on the data from Table 1(a). The differences between the inferences of some methods and others are small since the sample size $n=871$ is very high. These differences are noticed slightly more in the case of the negative predictive values, since they are more different to each other than the positive ones: $\hat{P}_A = 0.8935$, $\hat{P}_B =$

0.8807, $\hat{N}_A = 0.7849$ and $\hat{N}_B = 0.6478$.

It should be reminded the recommendation made by Roldán et al.[2] to carry out the global homogeneity test before performing an individual analysis. For the example data they use ($x_1$=152, $x_2$=17, $x_2$=7, $x_4$=36, $x_5$=25, $x_6$=10, $x_7$=11, and $x_8$=290), the classic global test provides the value $\chi_d^2$=4.2517 (*p-value*=0.1193) and the two tests selected -methods *R* and *LR*- provide the values $\chi_R^2$=4.2067 (*p-value*=0.1220) and $\chi_{LR}^2$=4.2064 (*p-value*=0.1221). None of the tests is significant, but the individual test for the negative predictive values is always significant: $z_d^2$=4.2461 (*p-value*=0.0393), $z_{d(p)}^2$=4.2076 (*p-value*=0.0402) and $z_{d(a)}^2$=4.0730 (*p-value*=0.0445).

## 8. Conclusions

The inferences about the difference $d$ ($\bar{d}$) or the ratio $R$ ($\bar{R}$) of the positive (negative) predictive values of two diagnostic tests *A* and *B* (paired samples) were not very well developed until the year 2006[9]. These inferences may refer to confidence intervals, confidence regions, individual homogeneity tests, tests of non-inferiority, or global homogeneity tests. Over the last few years, there have essentially been four procedures that have emerged -those of Wang et al.[7], Moskowitz and Pepe[9], Roldán et al.[2] and Kosinski[8]-, each one devoted to one or more types of inference. In all cases, different expressions are provided for the inferences about the positive and negative predictive values; this article has shown that this can be avoided applying the two properties of symmetry from Section 2. In all cases, complex expressions are used, except partially in the case of the last author. In this article, more simple and intuitive expressions have been used. Furthermore, the inferences for *R* are based on the logarithm of its $\hat{R}$ estimator more than on $\hat{R}$ itself; this article also proposes inferences based on $\hat{R}$ itself. Moreover, the inference methods proposed are modified in two senses: (1) obtaining from them a Wald-type adjusted method, which consists of adding 0.5 to all of the frequencies observed, which provides the test methods or the *CI* methods *d(a)*, *LR(a)* and *R(a)*; or (2) obtaining from them a pooled-type method[8], which consists of estimating the predictive values that appear in the formulae of the variance or covariance under the assumption of their homogeneity, which provides the test methods *d(p)*, *LR(p)* and *R(p)*.

For three of the inferences -confidence intervals and individual or global homogeneity test- different simulations were performed to select the best inference method. In the case of the *CI* for the difference parameter, it is observed that the method of Wang *et al.*[7] may have a very low coverage, and this improves considerably if the data increase by 0.5 i.e. with the *d(a)* method or adjusted-Wang *et al* method. In the case of the *CI* for the ratio parameter, it is observed that we should not use the methods *LR* by Wang *et al.*[7] and Moskowitz and Pepe[9], nor the current *R*, since they may also have a very low coverage, and both improve considerably if the data increase by 0.5, i.e. with the *LR(a)* and *R(a)* methods. Both methods perform in a similar way, although the *R(a)* method is a little better.

In the case of the individual homogeneity test of the positive (or negative) predictive values, it once again happens that we should not use the classic methods *d* or *LR* by Wang *et al.*[7] and Moskowitz and Pepe[9], nor the current *R*, since all of them may be very liberal. In this case, the best choice is the *d(p)* method by Kosinski[8], followed very closely by the new *R(p)* method, which is much more powerful than the previous one if the prevalences are low and the differences of the positive and negative predictive values are equal.

Finally, in the case of the global homogeneity test of the positive and negative predictive values, respectively, it is observed that we should not use the classic *d* method used by Roldán *et al.*[2], since this method is always liberal. The other eight methods assessed in this article verify the opposite -they are always conservative-, but method *R* always has the same or more power than the rest, and therefore it should be selected. Its differences in power with the rest of the valid methods are more notable the lower *n* is, and less notable when both predictive values are different, but in a different way. Nevertheless, method *LR* is only slightly worse than method *R*.

## 9. Acknowledgements

This research was supported by the Ministry of Science and Innovation (Spain), Grant PID2021-126095NB-I00 funded by MCIN/AEI/10.13039/501100011033 and by "ERDF A way of making Europe".

**APPENDIX A. Covariance matrix of the vectors ($d$, $\bar{d}$), ($\log R$, $\log \bar{R}$), and ($R$, $\bar{R}$)**

The demonstrations made by different authors -Wang *et al.*[7] and Roldán *et al.*[2]- have two practical difficulties: they use matrix-type demonstrations and they do not substitute the values obtained by the predictive values that are equivalent to them. Modifying these two aspects, more simple and intuitive expressions are obtained, which are also more advantageous to determine the right sample size. In the following, it is necessary to take into account that if $\hat{p}_i = x_i/n$, $\boldsymbol{p} = (p_1, p_2, \ldots, p_8)$ and $\hat{\boldsymbol{p}} = (\hat{p}_1, \hat{p}_2, \ldots, \hat{p}_8)$, then $\hat{\boldsymbol{p}} - \boldsymbol{p} \to N(\boldsymbol{0}, \Sigma)$, where $\Sigma = (\sigma_{ij})$, $\sigma_{ii} = p_i(1-p_i)/n$, and $\sigma_{ij} = -p_i p_j/n$ (if $i \neq j$). Therefore, given any function $f(\boldsymbol{p})$, estimated by $\hat{f} = f(\hat{\boldsymbol{p}})$ and with derivatives $f_i = \partial f/\partial p_i$, the delta method[18] indicates that $V(\hat{f}) = \Sigma f_i^2 \sigma_{ii} - \Sigma_i \Sigma_{j \neq i} f_i f_j \sigma_{ij} = [\Sigma f_i^2 p_i - (\Sigma f_i p_i)^2]/n$. Something similar is obtained for the covariance between two functions $\hat{f}$ and $\hat{g}$, with $\hat{g}(\hat{\boldsymbol{p}})$ the estimation of $g(\boldsymbol{p})$ and $g_i = \partial g/\partial p_i$. With this

$$nV(\hat{f}) = \Sigma f_i^2 p_i - (\Sigma f_i p_i)^2 \text{ and } n\text{Cov}(\hat{f}, \hat{g}) = \Sigma f_i g_i p_i - (\Sigma f_i p_i)(\Sigma g_i p_i). \tag{A1}$$

The estimated values $n\hat{V}(\hat{f})$ and $n\widehat{\text{Cov}}(\hat{f}, \hat{g})$ are obtained substituting in the two previous expressions $p_i$ with $\hat{p}_i$. In the following, it is necessary to take into account that in the current case: $(p_1+p_2) = t_A P_A$, $(p_1+p_3) = t_B P_B$, $(p_5+p_6) = t_A(1-P_A)$ and $(p_5+p_7) = t_B(1-P_B)$.

For the case ($d$, $\bar{d}$) we have that $f = d = (p_1+p_2)/(t_1+t_2) - (p_1+p_3)/(t_1+t_3)$, $g = \bar{d} =$

$(p_7+p_8)/(t_3+t_4)-(p_6+p_8)/(t_2+t_4)$, $\hat{f}=\hat{d}$, and $\hat{g}=\hat{\bar{d}}$, and therefore $f_1=f_2+f_3$, $f_2=(1-P_A)/t_A$, $f_3= -(1-P_B)/t_B$, $f_4=f_8=0$, $f_5=f_6+f_7$, $f_6=-P_A/t_A$ and $f_7=P_B/t_B$. The values of $g_i$ are obtained applying the property of symmetry in diagnoses: $g_8=g_6+g_7$, $g_7=(1-N_A)/\bar{t}_A$, $g_6=-(1-N_B)/\bar{t}_B$, $g_1=g_5=0$, $g_4=g_2+g_3$, $g_3=-N_A/\bar{t}_A$, and $g_2=N_B/\bar{t}_B$. As $\Sigma f_i p_i=\Sigma g_i p_i=0$ and some values $f_i$ and $g_i$ also take the value 0, the expressions (A1) are simplified leading to the following values of V($\hat{d}$)=$\sigma_d^2$, V($\hat{\bar{d}}$)= $\sigma_{\bar{d}}^2$, and Cov ($\hat{d},\hat{\bar{d}}$)=$\sigma_{d\bar{d}}$,

$$\sigma_d^2 = \frac{P_A(1-P_A)}{nt_A} + \frac{P_B(1-P_B)}{nt_B} - 2\frac{(1-P_A)(1-P_B)p_1 + P_A P_B p_5}{nt_A t_B},$$

$$\sigma_{\bar{d}}^2 = \frac{N_A(1-N_A)}{n\bar{t}_A} + \frac{N_B(1-N_B)}{n\bar{t}_B} - 2\frac{(1-N_A)(1-N_B)p_8 + N_A N_B p_4}{n\bar{t}_A \bar{t}_B}, \text{ and}$$

$$\sigma_{d\bar{d}} = \frac{(1-P_A)N_B p_2 + P_A(1-N_B)p_6}{nt_A \bar{t}_B} + \frac{(1-P_B)N_A p_3 + P_B(1-N_A)p_7}{n\bar{t}_A t_B}.$$

Substituting each value with its estimation, expressions (1) to (3) are obtained.

Moskowitz and Pepe[9] point out that the covariance matrix (log $R$, log $\bar{R}$) is very difficult to obtain, and therefore they determine it through multinomial-Poisson transformation. As can be seen now, there are not so many difficulties. Now $f=\log R= \log(p_1+p_2)-\log(t_1+t_2)+\log(p_1+p_3)-\log(t_1+t_3)$, $g=\log \bar{R} =\log(p_7+p_8)-\log(t_3+t_4)+\log(p_6+p_8)-\log(t_2+t_4)$, $\hat{f}=\log\hat{R}$, and $\hat{g}=\log\hat{\bar{R}}$. With this $f_1=f_2+f_3$, $f_2=(1-P_A)/P_A t_A$, $f_3=-(1-P_B)/P_B t_B$, $f_4=f_8=0$, $f_5=f_6+f_7$, $f_6=-1/t_A$, and $f_7=-1/t_B$. The values of $g_i$ are obtained applying the property of symmetry in diagnoses: $g_8=g_6+g_7$, $g_7=(1-N_A)/N_A \bar{t}_A$, $g_6=-(1-N_B)/N_B \bar{t}_B$, $g_1=g_5=0$, $g_4=g_2+g_3$, $g_3=-1/\bar{t}_A$, and $g_2=-1/\bar{t}_B$. Applying the expressions (A1) we obtain the following values of V(log $\hat{R}$)=$\sigma_R^2$, V(log $\hat{\bar{R}}$)=$\sigma_{\bar{R}}^2$ and Cov(log $\hat{R}$, log $\hat{\bar{R}}$)=$\sigma_{R\bar{R}}$,

$$\sigma_R^2 = \frac{1-P_A}{nt_A P_A} + \frac{1-P_B}{nt_B P_B} - \frac{2}{nt_A t_B}\left\{\frac{(1-P_A)(1-P_B)}{P_A P_B}p_1 + p_5\right\},$$

$$\sigma_{\bar{R}}^2 = \frac{1-N_A}{n\bar{t}_A N_A} + \frac{1-N_B}{n\bar{t}_B N_B} - \frac{2}{n\bar{t}_A \bar{t}_B} \left\{ \frac{(1-N_A)(1-N_B)}{N_A N_B} p_8 + p_4 \right\} \text{ and}$$

$$\sigma_{R\bar{R}} = \frac{1}{nt_A \bar{t}_B} \left\{ \frac{1-P_A}{P_A} p_2 + \frac{1-N_B}{N_B} p_6 \right\} + \frac{1}{n\bar{t}_A t_B} \left\{ \frac{1-P_B}{P_B} p_3 + \frac{1-N_A}{N_A} p_7 \right\}.$$

Substituting each value with its estimation, expressions (6) to (8) are obtained.

Moreover, the traditional method of obtaining the approximate distribution of a relative risk $\hat{R}$ consists of applying the delta method to the function $\hat{f} = \log \hat{R}$. As $V(\log \hat{R}) = V(\hat{R})/R^2$, then $V(\hat{R}) = R^2 V(\log \hat{R})$. Applying that expression to the current case it is obtained that $V(\hat{R}) = R^2 \sigma_R^2$. With this, the test statistic for $H_{R(\rho)}$ will be $(\hat{R}-\rho)/\sqrt{R^2 \sigma_R^2}$, where $\sigma_R^2$ must be substituted with its estimator $\hat{\sigma}_R^2$ and $R^2$ can be substituted by: *(i)* its unrestricted estimator $\hat{R}^2$; or *(ii)* its $\rho^2$ value under $H_{R(\rho)}$; or *(iii)* the $\rho\hat{R}$ value, which is a mix of the two previous procedures. As only the last procedure respects the property of symmetry in the tests, then the statistic will be $z = (\hat{R}-\rho)/\sqrt{\rho \hat{R} \sigma_R^2}$. A similar reasoning applies for $\hat{g} = \log \hat{\bar{R}}$. This explains the results of the individual inferences of expressions (11), in which the $CI_R$ is obtained solving in $\rho$ the equality $z_\alpha^2 = (\hat{R}-\rho)^2 / (\rho \hat{R} \hat{\sigma}_R^2)$. The global inferences of expressions (12) are deduced from the fact that $\text{Cov}(\hat{f}, \hat{g}) = \text{Cov}(\hat{R}, \hat{\bar{R}})/R\bar{R} = \sigma_{R\bar{R}}/R\bar{R}$ so that $\widehat{\text{Cov}}(\hat{R}, \hat{\bar{R}}) = \hat{\sigma}_{R\bar{R}} \sqrt{\hat{R}\hat{\bar{R}}\rho\bar{\rho}}$ for the same reason as before.

Finally, the expressions of $\sigma_d^2$, $\sigma_{\bar{d}}^2$, $\sigma_R^2$ and $\sigma_{\bar{R}}^2$ are key in the formulae of the sample size of Wang *et al.*[7] and Moskowitz and Pepe[9], but the current format -which is very different to that of the aforementioned authors- allows us to work more easily with their formulae. For example, in the one-tailed test for the null hypotheses $H_{d(\delta)}$ or $H_{R(\rho)}$ to an error $\alpha$, and for a power $1-\beta$ in the alternative hypotheses $\delta_1 > \delta$ or $\rho_1 > \rho$, the sample sizes *n* are:

$$n = \left(\frac{z_{2\alpha}+z_{2\beta}}{\delta-\delta_1}\right)^2 \left\{\frac{P_A(1-P_A)}{t_A} + \frac{P_B(1-P_B)}{t_B} - 2\frac{(1-P_A)(1-P_B)p_1 + P_A P_B p_5}{t_A t_B}\right\} \text{ and}$$

$$n = \left(\frac{z_{2\alpha}+z_{2\beta}}{\log\rho-\log\rho_1}\right)^2 \left[\frac{1-P_A}{t_A P_A} + \frac{1-P_B}{t_B P_B} - \frac{2}{t_A t_B}\left\{\frac{(1-P_A)(1-P_B)}{P_A P_B}p_1 + p_5\right\}\right],$$

respectively. Note that the value of $n$ depends on the (unknown) value of the following parameters: *(i)* the predictive values $P_A$ and $P_B$ of tests $A$ and $B$, respectively; *(ii)* the proportions of positive diagnoses $t_A$ and $t_B$ of tests $A$ and $B$, respectively; and *(iii)* the proportions $p_1$ and $p_5$, i.e. the proportions of diagnoses $(A, B, S)=(+, +. +)$ and $(+, +. -)$, respectively.

**APPENDIX B. Other statistics from the Bennett function**

Bennett[14] justified the idea that the null hypothesis of equality of positive predictive values is equivalent to the null hypothesis $H'$: $f'=(p_2-p_3)/(p_1+p_2+p_3+p_4)-\{(p_1+p_2)(p_6+p_8)-(p_1+p_3)(p_7+p_8)\}/\{(p_1+p_2+p_3+p_4)(p_5+p_6+p_7+p_8)\}=0$, when $f'$ is the Bennett function which Wu[15] argues about. Carrying out some operations, we obtain the new null hypothesis $H$: $f=(p_1+p_2)(p_5+p_7)-(p_1+p_3)(p_5+p_6)=t_A t_B (P_A-P_B)=0$, where the function $f$ is estimated by:

$$\hat{f} = a/n \text{ where } a = (x_1+x_2)(x_5+x_7)-(x_1+x_3)(x_5+x_6) \quad \text{(B1)}$$

$$= n_A n_B (\hat{P}_A - \hat{P}_B)/n^2 \quad \text{(B2)}$$

This function $f$ is the same as that which is obtained by carrying out operations in the equality $P_A-P_B=0$ since $f=(p_1+p_2)(t_1+t_3)-(p_1+p_3)(t_1+t_2)$. Applying the delta method from the first paragraph in Appendix A, $f_1=p_7-p_6$, $f_2=p_5+p_7$, $f_3=-(p_5+p_6)$, $f_4=0$, $f_5=p_2-p_3$, $f_6=-(p_1+p_3)$, $f_7=p_1+p_2$, and $f_8=0$. Therefore,

$$M = \Sigma f_i p_i = 2f = 2\{(p_1+p_2)(p_5+p_7)-(p_1+p_3)(p_5+p_6)\} \quad \text{(B3)}$$

$$= 2t_A t_B (P_A-P_B), \quad \text{(B4)}$$

$$F = \Sigma f_i^2 p_i = p_1(p_6-p_7)^2 + p_2(p_5+p_7)^2 + p_3(p_5+p_6)^2 + p_7(p_1+p_2)^2 + p_6(p_1+p_3)^2 + p_5(p_2-p_3)^2 \quad \textbf{(B5)}$$

$$= t_A t_B [t_A P_B (1-P_B) + t_B P_A (1-P_A) + (t_A+t_B)(P_A-P_B)^2 - 2 \times \{p_1(1-P_A)(1-P_B) + p_5 P_A P_B\}], \quad \textbf{(B6)}$$

where the last expression is due to the fact that expression (B5) may also be written as $F = (p_1+p_2)(p_5+p_7)^2 + (p_1+p_3)(p_5+p_6)^2 + (p_5+p_6)(p_1+p_3)^2 + (p_5+p_7)(p_1+p_2)^2 - 2 \times [p_1(p_5+p_6)(p_5+p_7) + p_5(p_1+p_2)(p_1+p_3)]$. It should be noted that the previous expressions have been set in terms of the $p_i$ parameters and also in terms of the predictive values. The objective is that the final formulae have the two formats: the classic ones of Bennett[14] and Wu[15] and the more intuitive one by Kosinski[8]. The test statistic is $z_{exp}^2 = \hat{f}^2 / \hat{V}(\hat{f})$ where, through expression (A1), $\hat{V}(\hat{f})$ is the estimator of $V(\hat{f}) = \{\Sigma f_i^2 p_i - (\Sigma f_i p_i)^2\}/n = \{F - M^2\}/n$. But that estimation can be made subject to three criteria, depending on how $F$ and $M$ are estimated.

If $M$ is estimated under $H$ -i.e. assuming that $P_A = P_B = P$-, but $F$ is estimated in an unrestricted way -i.e. making $\hat{p}_i = x_i/n$-, then $M=0$ and the following statistic of Bennett[14] is obtained, with its errors already corrected. Using expressions (B1) and (B5)

$$z_B^2 = a^2 / (b_0 + b_1) \text{ where } a = (x_1+x_2)(x_5+x_7) - (x_1+x_3)(x_5+x_6),$$

where $b_0 = x_1(x_6-x_7)^2 + x_2(x_5+x_7)^2 + x_3(x_5+x_6)^2$ and $b_1 = x_7(x_1+x_2)^2 + x_6(x_1+x_3)^2 + x_5(x_2-x_3)^2$. Bennett[14] provided the correct value $b_0$, but he was wrong in the term $b_1$ because he put the trio ($x_6$, $x_7$, $x_8$) instead of the correct trio ($x_7$, $x_6$, $x_5$). In an alternative and equivalent way, if we use the expressions (B2) and (B6) then, for $\hat{\sigma}_d^2$ as in expression (1),

$$z_B^2 = (\hat{P}_A - \hat{P}_B)^2 / \left\{ \hat{\sigma}_d^2 + (\hat{P}_A - \hat{P}_B)^2 (1/n_A + 1/n_B) \right\}.$$

If the two parameters $F$ and $M$ are estimated in an unrestricted way then -using expressions (B1), (B3) and (B5)- the test statistic is

$$z_{B'}^2 = a^2 / \{b_0 + b_1 - 4a^2/n\}.$$

In an alternative and equivalent way, if we use expressions (B2), (B4), and (B6) then,

$$z_{B'}^2 = \left(\hat{P}_A - \hat{P}_B\right)^2 \Big/ \left\{\hat{\sigma}_d^2 + \left(\hat{P}_A - \hat{P}_B\right)^2 \left(1/n_A + 1/n_B - 4/n\right)\right\}.$$

As Wu[15] is based on function $f'$ instead of function $f$, its statistic $z_W^2$ differs from the previous one $z_{B'}^2$ in number 4 must be substituted by $\Pi = \left\{\hat{\pi}^{-1} + (1-\hat{\pi})^{-1} - 1\right\}$, where $\hat{\pi} = (x_1 + x_2 + x_3 + x_4)/n$ is the estimator of prevalence. The value $\Pi$ is very different to 4 when $\hat{\pi}$ is very near to 0 or 1.

Finally, if $F$ and $M$ are estimated assuming that $P_A = P_B = P$ then -using the last expressions of (B2), (B4) and (B6)- we obtain the statistic $z_{d(p)}^2$ of Kosinski[8].

In all cases, the test for $\bar{H}: N_A = N_B$ is obtained applying the property of symmetric in diagnoses (in this case, the formulae of Bennett have more mistakes than in the previous case). For the data from the example from Table 1(a): $z_B^2 = 0.800$ and $z_{B'}^2 = z_W^2 = 0.803$ in the case of hypothesis $P_A = P_B$, and $z_B^2 = 20.220$, $z_{B'}^2 = 22.290$ and $z_W^2 = 22.145$ in the case of hypothesis $N_A = N_B$. The authors have observed that that statistic $z_{B'}^2$ does not improve the statistics selected in this article.

## Table 1

**Paired study design ($S$ = gold standard; $A$ and $B$ = tests to be compared)**

**(a) Notation and numerical values of the example CAD (coronary artery disease) by Weiner *et al.* (1979), where $A$ = "result of clinical history" and $B$ = "result of stress test".**

| $A$ | + | | − | | |
|---|---|---|---|---|---|
| $B$ | + | − | + | − | Total |
| $S = +$ | $x_1 = 473$ | $x_2 = 81$ | $x_3 = 29$ | $x_4 = 25$ | |
| $S = -$ | $x_5 = 22$ | $x_6 = 44$ | $x_7 = 46$ | $x_8 = 151$ | |
| Totals | $n_1 = 495$ | $n_2 = 125$ | $n_3 = 75$ | $n_4 = 176$ | $n = 871$ |

Extra notation:    $n_A = n_1+n_2$,    $n_B = n_1+n_3$,    $\bar{n}_A = n_3+n_4$,    $\bar{n}_B = n_2+n_4$

$x_A = x_1+x_2$,    $x_B = x_1+x_3$,    $\bar{x}_A = x_7+x_8$,    $\bar{x}_B = x_6+x_8$

**(b) Probability of obtaining an observation in each one of the cells in Table (a) under the multinomial model**

| $A$ | + | | − | | |
|---|---|---|---|---|---|
| $B$ | + | − | + | − | Total |
| $S = +$ | $p_1$ | $p_2$ | $p_3$ | $p_4$ | |
| $S = -$ | $p_5$ | $p_6$ | $p_7$ | $p_8$ | |
| Totals | $t_1$ | $t_2$ | $t_3$ | $t_4$ | 1 |

Extra notation:    $t_A = t_1+t_2$,    $t_B = t_1+t_3$,    $\bar{t}_A = t_3+t_4$,    $\bar{t}_B = t_2+t_4$

$p_A = p_1+p_2$,    $p_B = p_1+p_3$,    $\bar{p}_A = p_7+p_8$,    $\bar{p}_B = p_6+p_8$

**Table 2**

**Empirical coverage (C, in %) and average width (W) of the 95%-CIs that are obtained with the six inference methods that are indicated**

| $P_A$ | $P_B$ | $N_A$ | $N_B$ | $\pi$ | $O^+$ | $O^-$ | n | Parameter | | \multicolumn{12}{c}{Inference Method} | | | | | | | | | | | |
|---|---|---|---|---|---|---|---|---|---|---|---|---|---|---|---|---|---|---|---|---|
| | | | | | | | | | | d | | d(a) | | LR | | LR(a) | | R | | R(a) | |
| | | | | | | | | d | R | C | W | C | W | C | W | C | W | C | W | C | W |
| .8 | .8 | .8 | .8 | .35 | 5 | 2 | 100 | 0 | 1.00 | 93.9 | .376 | 95.6 | .364 | 95.6 | 0.498 | 96.8 | 0.491 | 95.5 | 0.496 | 96.8 | 0.490 |
| | | | | | | | 200 | 0 | 1.00 | 94.5 | .269 | 95.2 | .264 | 95.3 | 0.345 | 95.9 | 0.343 | 95.2 | 0.344 | 95.9 | 0.342 |
| | | | | | | | 300 | 0 | 1.00 | 94.6 | .220 | 95.1 | .217 | 95.2 | 0.280 | 95.6 | 0.279 | 95.2 | 0.280 | 95.6 | 0.278 |
| | | | | | 2 | 5 | 100 | 0 | 1.00 | 94.1 | .358 | 95.9 | .350 | 95.8 | 0.473 | 97.2 | 0.472 | 95.8 | 0.472 | 97.2 | 0.471 |
| | | | | | | | 200 | 0 | 1.00 | 94.5 | .255 | 95.4 | .252 | 95.4 | 0.327 | 96.1 | 0.327 | 95.4 | 0.327 | 96.1 | 0.327 |
| | | | | | | | 300 | 0 | 1.00 | 94.7 | .209 | 95.2 | .207 | 95.3 | 0.266 | 95.7 | 0.265 | 95.2 | 0.265 | 95.7 | 0.265 |
| | | .8 | .7 | .35 | 5 | 2 | 100 | 0 | 1.00 | 90.0 | .519 | 96.6 | .504 | 92.8 | 2.636 | 98.6 | 0.808 | 92.7 | 0.961 | 98.5 | 0.800 |
| | | | | | | | 200 | 0 | 1.00 | 92.9 | .374 | 95.4 | .367 | 94.2 | 0.511 | 96.7 | 0.519 | 94.1 | 0.509 | 96.7 | 0.518 |
| | | | | | | | 300 | 0 | 1.00 | 93.6 | .308 | 95.2 | .303 | 94.5 | 0.406 | 96.1 | 0.412 | 94.4 | 0.405 | 96.1 | 0.411 |
| | | | | | 2 | 5 | 100 | 0 | 1.00 | 90.6 | .507 | 96.8 | .497 | 93.5 | 2.591 | 98.8 | 0.796 | 93.4 | 0.944 | 98.7 | 0.789 |
| | | | | | | | 200 | 0 | 1.00 | 92.9 | .365 | 95.6 | .360 | 94.2 | 0.497 | 96.9 | 0.508 | 94.2 | 0.495 | 96.9 | 0.507 |
| | | | | | | | 300 | 0 | 1.00 | 93.7 | .299 | 95.3 | .296 | 94.5 | 0.395 | 96.3 | 0.402 | 94.5 | 0.394 | 96.3 | 0.401 |
| .8 | .8 | .8 | .8 | .65 | 5 | 2 | 100 | 0 | 1.00 | 94.7 | .159 | 95.6 | .159 | 95.2 | 0.201 | 96.0 | 0.203 | 95.2 | 0.201 | 96.0 | 0.203 |
| | | | | | | | 200 | 0 | 1.00 | 94.9 | .113 | 95.3 | .113 | 95.1 | 0.142 | 95.5 | 0.143 | 95.1 | 0.142 | 95.5 | 0.143 |
| | | | | | | | 300 | 0 | 1.00 | 94.9 | .092 | 95.2 | .092 | 95.1 | 0.116 | 95.3 | 0.116 | 95.1 | 0.116 | 95.3 | 0.116 |
| | | | | | 2 | 5 | 100 | 0 | 1.00 | 94.8 | .138 | 96.0 | .140 | 95.4 | 0.175 | 96.4 | 0.179 | 95.3 | 0.175 | 96.4 | 0.179 |
| | | | | | | | 200 | 0 | 1.00 | 94.9 | .098 | 95.5 | .099 | 95.1 | 0.124 | 95.7 | 0.125 | 95.2 | 0.124 | 95.7 | 0.125 |
| | | | | | | | 300 | 0 | 1.00 | 94.9 | .080 | 95.3 | .081 | 95.1 | 0.101 | 95.4 | 0.102 | 95.1 | 0.101 | 95.5 | 0.102 |
| | | .8 | .7 | .65 | 5 | 2 | 100 | 0 | 1.00 | 94.7 | .165 | 95.6 | .165 | 95.2 | 0.209 | 96.1 | 0.211 | 95.2 | 0.209 | 96.0 | 0.211 |
| | | | | | | | 200 | 0 | 1.00 | 94.9 | .117 | 95.3 | .117 | 95.1 | 0.147 | 95.5 | 0.148 | 95.1 | 0.147 | 95.5 | 0.148 |
| | | | | | | | 300 | 0 | 1.00 | 94.9 | .096 | 95.2 | .096 | 95.1 | 0.120 | 95.4 | 0.121 | 95.1 | 0.120 | 95.3 | 0.121 |
| | | | | | 2 | 5 | 100 | 0 | 1.00 | 94.8 | .144 | 95.9 | .146 | 95.4 | 0.183 | 96.4 | 0.187 | 95.4 | 0.183 | 96.4 | 0.187 |
| | | | | | | | 200 | 0 | 1.00 | 94.9 | .103 | 95.5 | .103 | 95.2 | 0.129 | 95.7 | 0.131 | 95.2 | 0.129 | 95.7 | 0.131 |
| | | | | | | | 300 | 0 | 1.00 | 94.9 | .084 | 95.3 | .084 | 95.1 | 0.105 | 95.5 | 0.106 | 95.1 | 0.105 | 95.5 | 0.106 |
| .8 | .7 | .8 | .8 | .35 | 5 | 2 | 100 | .1 | 1.14 | 94.0 | .373 | 95.2 | .360 | 95.4 | 0.613 | 96.1 | 0.593 | 95.3 | 0.611 | 96.0 | 0.592 |
| | | | | | | | 200 | .1 | 1.14 | 94.5 | .265 | 95.1 | .260 | 95.2 | 0.422 | 95.5 | 0.415 | 95.2 | 0.421 | 95.5 | 0.415 |
| | | | | | | | 300 | .1 | 1.14 | 94.7 | .217 | 95.1 | .214 | 95.1 | 0.341 | 95.4 | 0.338 | 95.1 | 0.341 | 95.4 | 0.338 |
| | | | | | 2 | 5 | 100 | .1 | 1.14 | 94.2 | .354 | 95.4 | .345 | 95.4 | 0.583 | 96.2 | 0.569 | 95.4 | 0.582 | 96.1 | 0.568 |

| $P_A$ | $P_B$ | $N_A$ | $N_B$ | $\pi$ | $O^+$ | $O^-$ | n | Parameter | | \multicolumn{12}{c}{Inference Method} |
|---|---|---|---|---|---|---|---|---|---|---|---|---|---|---|---|---|---|---|---|---|
| | | | | | | | | | | d | | d(a) | | LR | | LR(a) | | R | | R(a) | |
| | | | | | | | | d | R | C | W | C | W | C | W | C | W | C | W | C | W |
| .8 | .7 | .8 | .8 | .35 | 2 | 5 | 200 | .1 | 1.14 | 94.6 | .251 | 95.2 | .248 | 95.2 | 0.400 | 95.6 | 0.396 | 95.2 | 0.400 | 95.6 | 0.395 |
| | | | | | | | 300 | .1 | 1.14 | 94.7 | .206 | 95.1 | .204 | 95.2 | 0.324 | 95.4 | 0.322 | 95.1 | 0.324 | 95.4 | 0.321 |
| | | .8 | .7 | .35 | 5 | 2 | 100 | .1 | 1.14 | 92.0 | .523 | 95.1 | .492 | 93.5 | 2.476 | 96.2 | 0.969 | 93.4 | 1.148 | 96.1 | 0.959 |
| | | | | | | | 200 | .1 | 1.14 | 93.7 | .375 | 94.9 | .362 | 94.3 | 0.651 | 95.7 | 0.639 | 94.3 | 0.649 | 95.6 | 0.637 |
| | | | | | | | 300 | .1 | 1.14 | 94.2 | .307 | 94.9 | .300 | 94.6 | 0.516 | 95.4 | 0.511 | 94.5 | 0.515 | 95.4 | 0.510 |
| | | | | | 2 | 5 | 100 | .1 | 1.14 | 92.0 | .508 | 95.3 | .481 | 93.4 | 2.493 | 96.5 | 0.949 | 93.3 | 1.130 | 96.4 | 0.940 |
| | | | | | | | 200 | .1 | 1.14 | 93.7 | .363 | 95.0 | .352 | 94.3 | 0.633 | 95.8 | 0.623 | 94.2 | 0.630 | 95.7 | 0.621 |
| | | | | | | | 300 | .1 | 1.14 | 94.1 | .297 | 95.0 | .291 | 94.6 | 0.501 | 95.5 | 0.497 | 94.5 | 0.500 | 95.5 | 0.496 |
| .8 | .7 | .8 | .8 | .65 | 5 | 2 | 100 | .1 | 1.14 | 94.5 | .142 | 94.9 | .143 | 94.6 | 0.226 | 95.0 | 0.228 | 94.6 | 0.226 | 95.0 | 0.228 |
| | | | | | | | 200 | .1 | 1.14 | 94.7 | .101 | 95.0 | .101 | 94.8 | 0.159 | 95.0 | 0.160 | 94.8 | 0.159 | 95.0 | 0.160 |
| | | | | | | | 300 | .1 | 1.14 | 94.8 | .083 | 95.0 | .083 | 94.9 | 0.130 | 95.0 | 0.130 | 94.9 | 0.130 | 95.0 | 0.130 |
| | | | | | 2 | 5 | 100 | .1 | 1.14 | 94.2 | .133 | 94.7 | .134 | 94.3 | 0.212 | 94.7 | 0.214 | 94.3 | 0.212 | 94.7 | 0.214 |
| | | | | | | | 200 | .1 | 1.14 | 94.6 | .094 | 94.9 | .095 | 94.6 | 0.149 | 94.8 | 0.150 | 94.6 | 0.149 | 94.8 | 0.150 |
| | | | | | | | 300 | .1 | 1.14 | 94.8 | .077 | 94.9 | .077 | 94.8 | 0.122 | 94.9 | 0.122 | 94.8 | 0.122 | 94.9 | 0.122 |
| | | .8 | .7 | .65 | 5 | 2 | 100 | .1 | 1.14 | 94.5 | .145 | 95.0 | .145 | 94.6 | 0.231 | 95.0 | 0.232 | 94.6 | 0.231 | 95.0 | 0.232 |
| | | | | | | | 200 | .1 | 1.14 | 94.7 | .103 | 95.0 | .103 | 94.8 | 0.163 | 95.0 | 0.163 | 94.8 | 0.163 | 95.0 | 0.163 |
| | | | | | | | 300 | .1 | 1.14 | 94.8 | .084 | 95.0 | .084 | 94.9 | 0.133 | 95.0 | 0.133 | 94.9 | 0.133 | 95.0 | 0.133 |
| | | | | | 2 | 5 | 100 | .1 | 1.14 | 94.3 | .135 | 94.8 | .136 | 94.3 | 0.216 | 94.8 | 0.218 | 94.3 | 0.215 | 94.8 | 0.218 |
| | | | | | | | 200 | .1 | 1.14 | 94.6 | .096 | 94.9 | .096 | 94.7 | 0.152 | 94.9 | 0.152 | 94.7 | 0.152 | 94.9 | 0.152 |
| | | | | | | | 300 | .1 | 1.14 | 94.8 | .078 | 94.9 | .078 | 94.8 | 0.124 | 94.9 | 0.124 | 94.8 | 0.124 | 94.9 | 0.124 |
| | | | | | | | | | Minimum | 90.0 | .077 | 94.7 | .077 | 92.8 | 0.101 | 94.7 | 0.102 | 92.7 | 0.101 | 94.7 | 92.8 |
| | | | | | | | | | Maximum | 94.9 | .518 | 96.8 | .504 | 95.8 | 2.636 | 98.8 | 0.969 | 95.8 | 1.148 | 98.7 | 95.8 |
| | | | | | | | | | Average | 94.6 | .185 | 95.2 | .220 | 94.8 | 0.478 | 95.8 | 0.338 | 94.8 | 0.352 | 95.8 | 94.8 |

**Table 3**

**Empirical size (in %) for the nine individual homogeneity tests of positive predictive values that are indicated (nominal error =5%)**

| $P_A$ | $P_B$ | $N_A$ | $N_B$ | $\pi$ | $O^+$ | $O^-$ | n | \multicolumn{9}{c}{*Inference Method*} |
|---|---|---|---|---|---|---|---|---|---|---|---|---|---|---|---|---|
| | | | | | | | | d | d(a) | d(p) | LR | LR(a) | LR(p) | R | R(a) | R(p) |
| .8 | .8 | .8 | .8 | .35 | 5 | 2 | 100 | 6.1 | 4.4 | 4.9 | 4.4 | 3.2 | 4.7 | 4.5 | 3.2 | 4.9 |
| | | | | | | | 200 | 5.6 | 4.8 | 5.0 | 4.7 | 4.1 | 4.9 | 4.8 | 4.1 | 5.0 |
| | | | | | | | 300 | 5.4 | 4.9 | 5.0 | 4.8 | 4.4 | 4.9 | 4.9 | 4.4 | 5.0 |
| | | | | | 2 | 5 | 100 | 5.9 | 4.1 | 4.8 | 4.2 | 2.8 | 4.7 | 4.2 | 2.9 | 4.8 |
| | | | | | | | 200 | 5.5 | 4.7 | 5.0 | 4.6 | 3.9 | 4.9 | 4.7 | 3.9 | 5.0 |
| | | | | | | | 300 | 5.3 | 4.8 | 5.0 | 4.8 | 4.3 | 5.0 | 4.8 | 4.3 | 5.0 |
| | | .8 | .7 | .35 | 5 | 2 | 100 | 10.0 | 3.4 | 4.2 | 7.2 | 1.4 | 4.9 | 7.2 | 1.5 | 5.0 |
| | | | | | | | 200 | 7.2 | 4.6 | 4.8 | 5.8 | 3.3 | 4.9 | 6.1 | 3.3 | 5.0 |
| | | | | | | | 300 | 6.4 | 4.8 | 4.9 | 5.6 | 3.9 | 5.0 | 5.7 | 3.9 | 5.0 |
| | | | | | 2 | 5 | 100 | 9.3 | 3.2 | 4.3 | 6.5 | 1.2 | 5.2 | 6.5 | 1.3 | 5.3 |
| | | | | | | | 200 | 7.1 | 4.4 | 4.8 | 5.8 | 3.1 | 5.0 | 5.9 | 3.1 | 5.0 |
| | | | | | | | 300 | 6.4 | 4.7 | 4.9 | 5.5 | 3.7 | 5.0 | 5.6 | 3.7 | 5.1 |
| .8 | .8 | .8 | .8 | .65 | 5 | 2 | 100 | 5.3 | 4.4 | 4.9 | 4.8 | 4.0 | 4.9 | 4.8 | 4.0 | 4.9 |
| | | | | | | | 200 | 5.2 | 4.7 | 5.0 | 4.9 | 4.5 | 5.0 | 4.9 | 4.5 | 5.0 |
| | | | | | | | 300 | 5.1 | 4.8 | 5.0 | 4.9 | 4.7 | 5.0 | 4.9 | 4.7 | 5.0 |
| | | | | | 2 | 5 | 100 | 5.2 | 4.1 | 4.9 | 4.7 | 3.6 | 4.8 | 4.7 | 3.6 | 4.8 |
| | | | | | | | 200 | 5.1 | 4.5 | 4.9 | 4.9 | 4.3 | 4.9 | 4.9 | 4.3 | 4.9 |
| | | | | | | | 300 | 5.1 | 4.7 | 5.0 | 4.9 | 4.6 | 5.0 | 4.9 | 4.6 | 5.0 |
| | | .8 | .7 | .65 | 5 | 2 | 100 | 5.3 | 4.4 | 4.9 | 4.8 | 4.0 | 4.9 | 4.8 | 4.0 | 4.9 |
| | | | | | | | 200 | 5.2 | 4.7 | 5.0 | 4.9 | 4.5 | 5.0 | 4.9 | 4.5 | 5.0 |
| | | | | | | | 300 | 5.1 | 4.8 | 5.0 | 4.9 | 4.7 | 5.0 | 4.9 | 4.7 | 5.0 |
| | | | | | 2 | 5 | 100 | 5.2 | 4.1 | 4.9 | 4.6 | 3.6 | 4.8 | 4.7 | 3.6 | 4.9 |
| | | | | | | | 200 | 5.1 | 4.6 | 4.9 | 4.8 | 4.3 | 4.9 | 4.9 | 4.3 | 4.9 |
| | | | | | | | 300 | 5.1 | 4.7 | 5.0 | 4.9 | 4.5 | 5.0 | 4.9 | 4.5 | 5.0 |
| | | | | | | | *Minimum* | 5.1 | 3.2 | 4.2 | 4.2 | 1.2 | 4.7 | 4.2 | 1.3 | 4.8 |
| | | | | | | | *Maximum* | 10.0 | 4.9 | 5.0 | 7.2 | 4.7 | 5.2 | 7.2 | 4.7 | 5.3 |
| | | | | | | | *Average* | 5.9 | 4.5 | 4.9 | 5.1 | 3.8 | 4.9 | 5.1 | 3.8 | 5.0 |

**Table 4**

**Empirical power that is obtained (in %) for the six individual homogeneity tests of positive predictive values that are indicated (nominal error $\alpha$=5%)**

| $P_A$ | $P_B$ | $N_A$ | $N_B$ | $\pi$ | $O^+$ | $O^-$ | $n$ | \multicolumn{6}{c}{Inference Method} | | | | | |
|---|---|---|---|---|---|---|---|---|---|---|---|---|---|
| | | | | | | | | d(a) | d(p) | LR(a) | LR(p) | R(a) | R(p) |
| .8 | .7 | .7 | .7 | .35 | 5 | 2 | 100 | 7.1 | 7.4 | 3.7 | 7.1 | 4.0 | 7.4 |
| | | | | | | | 200 | 12.7 | 12.7 | 10.2 | 11.9 | 10.4 | 12.3 |
| | | | | | | | 300 | 17.4 | 17.3 | 15.5 | 16.5 | 15.6 | 16.8 |
| | | | | | 2 | 5 | 100 | 6.8 | 7.2 | 3.3 | 7.2 | 3.5 | 7.5 |
| | | | | | | | 200 | 12.6 | 12.7 | 10.0 | 12.2 | 10.1 | 12.4 |
| | | | | | | | 300 | 17.5 | 17.4 | 15.5 | 16.8 | 15.6 | 17.0 |
| | | | | .65 | 5 | 2 | 100 | 71.7 | 72.8 | 71.1 | 72.3 | 71.1 | 72.5 |
| | | | | | | | 200 | 95.9 | 96.0 | 95.9 | 96.0 | 95.9 | 96.0 |
| | | | | | | | 300 | 99.6 | 99.6 | 99.6 | 99.6 | 99.6 | 99.6 |
| | | | | | 2 | 5 | 100 | 78.2 | 79.4 | 77.7 | 79.0 | 77.7 | 79.2 |
| | | | | | | | 200 | 98.1 | 98.2 | 98.1 | 98.1 | 98.1 | 98.2 |
| | | | | | | | 300 | 99.9 | 99.9 | 99.9 | 99.9 | 99.9 | 99.9 |
| .8 | .7 | .8 | .7 | .35 | 5 | 2 | 100 | 11.3 | 13.3 | 4.8 | 15.8 | 5.1 | 16.0 |
| | | | | | | | 200 | 18.4 | 20.4 | 12.5 | 22.6 | 12.7 | 22.8 |
| | | | | | | | 300 | 25.5 | 27.4 | 20.0 | 29.5 | 20.2 | 29.7 |
| | | | | | 2 | 5 | 100 | 11.4 | 14.3 | 4.5 | 16.8 | 4.9 | 17.1 |
| | | | | | | | 200 | 19.1 | 21.9 | 12.8 | 24.1 | 13.0 | 24.2 |
| | | | | | | | 300 | 26.7 | 29.3 | 20.9 | 31.4 | 21.0 | 31.5 |
| | | | | .65 | 5 | 2 | 100 | 76.7 | 78.1 | 75.9 | 77.9 | 75.9 | 78.0 |
| | | | | | | | 200 | 97.5 | 97.6 | 97.4 | 97.6 | 97.4 | 97.6 |
| | | | | | | | 300 | 99.8 | 99.8 | 99.8 | 99.8 | 99.8 | 99.8 |
| | | | | | 2 | 5 | 100 | 83.8 | 85.3 | 83.1 | 85.1 | 83.2 | 85.2 |
| | | | | | | | 200 | 99.1 | 99.2 | 99.1 | 99.1 | 99.1 | 99.1 |
| | | | | | | | 300 | 100.0 | 100.0 | 100.0 | 100.0 | 100.0 | 100.0 |
| .8 | .7 | .7 | .8 | .35 | 5 | 2 | 100 | 7.9 | 6.7 | 7.7 | 4.2 | 7.8 | 4.4 |
| | | | | | | | 200 | 15.4 | 13.7 | 16.3 | 10.2 | 16.4 | 10.5 |
| | | | | | | | 300 | 21.9 | 20.0 | 23.3 | 16.4 | 23.4 | 16.7 |
| | | | | | 2 | 5 | 100 | 7.5 | 6.2 | 7.2 | 4.0 | 7.3 | 4.1 |
| | | | | | | | 200 | 15.6 | 13.5 | 16.5 | 10.2 | 16.6 | 10.4 |
| | | | | | | | 300 | 22.5 | 20.3 | 24.0 | 16.7 | 24.1 | 16.9 |
| | | | | .65 | 5 | 2 | 100 | 72.8 | 73.9 | 72.4 | 73.2 | 72.5 | 73.5 |
| | | | | | | | 200 | 96.4 | 96.5 | 96.4 | 96.4 | 96.4 | 96.4 |
| | | | | | | | 300 | 99.6 | 99.6 | 99.6 | 99.6 | 99.6 | 99.6 |
| | | | | | 2 | 5 | 100 | 79.0 | 80.1 | 78.7 | 79.6 | 78.7 | 79.8 |
| | | | | | | | 200 | 98.3 | 98.4 | 98.3 | 98.3 | 98.3 | 98.3 |
| | | | | | | | 300 | 99.9 | 99.9 | 99.9 | 99.9 | 99.9 | 99.9 |
| | | | | | | | *Minimum* | 6.8 | 6.2 | 3.3 | 4.0 | 3.5 | 4.1 |
| | | | | | | | *Maximum* | 100.0 | 100.0 | 100.0 | 100.0 | 100.0 | 100.0 |
| | | | | | | | *Average* | 53.4 | 53.8 | 52.0 | 53.5 | 52.1 | 53.6 |

**Table 5**

**Empirical size that is obtained (in %) for the nine global homogeneity tests that are indicated (nominal error $\alpha$=5%)**

| $P_A$ | $P_B$ | $N_A$ | $N_B$ | $\pi$ | $O^+$ | $O^-$ | n | Inference Method | | | | | | | | |
|---|---|---|---|---|---|---|---|---|---|---|---|---|---|---|---|---|
| | | | | | | | | d | d(a) | d(p) | LR | LR(a) | LR(p) | R | R(a) | R(p) |
| .8 | .8 | .8 | .8 | .35 | 5 | 2 | 100 | 5.3 | 4.8 | 4.8 | 4.6 | 3.1 | 3.8 | 4.9 | 4.4 | 4.7 |
| | | | | | | | 200 | 5.9 | 4.1 | 4.0 | 4.9 | 4.1 | 4.5 | 4.6 | 3.1 | 3.9 |
| | | | | | | | 300 | 5.5 | 4.6 | 4.6 | 4.9 | 4.4 | 4.7 | 4.8 | 4.1 | 4.6 |
| | | | | | 2 | 5 | 100 | 5.3 | 4.8 | 4.8 | 4.5 | 3.1 | 3.9 | 4.9 | 4.4 | 4.7 |
| | | | | | | | 200 | 5.9 | 4.1 | 4.0 | 4.8 | 4.0 | 4.6 | 4.6 | 3.1 | 3.9 |
| | | | | | | | 300 | 5.5 | 4.6 | 4.6 | 4.9 | 4.4 | 4.7 | 4.9 | 4.1 | 4.6 |
| | | | | .65 | 5 | 2 | 100 | 5.3 | 4.8 | 4.7 | 4.5 | 3.1 | 3.9 | 4.9 | 4.4 | 4.7 |
| | | | | | | | 200 | 6.0 | 4.1 | 4.0 | 4.8 | 4.0 | 4.6 | 4.7 | 3.1 | 3.9 |
| | | | | | | | 300 | 5.5 | 4.6 | 4.6 | 4.9 | 4.4 | 4.7 | 4.9 | 4.1 | 4.5 |
| | | | | | 2 | 5 | 100 | 5.3 | 4.8 | 4.7 | 4.6 | 3.1 | 3.8 | 4.9 | 4.4 | 4.7 |
| | | | | | | | 200 | 6.7 | 2.2 | 2.2 | 4.9 | 4.1 | 4.5 | 4.0 | 1.5 | 2.3 |
| | | | | | | | 300 | 5.9 | 4.0 | 3.7 | 4.9 | 4.4 | 4.7 | 4.5 | 3.1 | 3.6 |
| .8 | .8 | .7 | .7 | .35 | 5 | 2 | 100 | 5.6 | 4.4 | 4.2 | 3.9 | 1.4 | 2.2 | 4.7 | 3.7 | 4.2 |
| | | | | | | | 200 | 6.3 | 2.2 | 2.3 | 4.4 | 3.0 | 3.6 | 3.9 | 1.5 | 2.6 |
| | | | | | | | 300 | 5.8 | 3.9 | 3.7 | 4.6 | 3.7 | 4.1 | 4.4 | 3.0 | 3.7 |
| | | | | | 2 | 5 | 100 | 5.6 | 4.3 | 4.3 | 3.8 | 1.5 | 2.5 | 4.6 | 3.7 | 4.3 |
| | | | | | | | 200 | 5.3 | 4.0 | 4.1 | 4.3 | 3.0 | 3.6 | 4.4 | 3.2 | 4.0 |
| | | | | | | | 300 | 5.2 | 4.6 | 4.5 | 4.6 | 3.6 | 4.2 | 4.7 | 4.1 | 4.5 |
| | | | | .65 | 5 | 2 | 100 | 5.1 | 4.7 | 4.7 | 4.4 | 3.2 | 3.9 | 4.8 | 4.4 | 4.7 |
| | | | | | | | 200 | 5.4 | 4.0 | 4.0 | 4.7 | 4.1 | 4.5 | 4.5 | 3.2 | 3.9 |
| | | | | | | | 300 | 5.2 | 4.5 | 4.5 | 4.8 | 4.4 | 4.7 | 4.8 | 4.1 | 4.5 |
| | | | | | 2 | 5 | 100 | 5.1 | 4.7 | 4.7 | 4.5 | 3.2 | 3.9 | 4.8 | 4.4 | 4.7 |
| | | | | | | | 200 | 5.3 | 4.8 | 4.8 | 4.7 | 4.1 | 4.4 | 4.9 | 4.4 | 4.7 |
| | | | | | | | 300 | 5.9 | 4.1 | 4.0 | 4.8 | 4.4 | 4.6 | 4.6 | 3.1 | 3.9 |
| | | | | | | | *Minimum* | 5.1 | 2.2 | 2.2 | 3.8 | 1.4 | 2.2 | 3.9 | 1.5 | 2.3 |
| | | | | | | | *Maximum* | 6.7 | 4.8 | 4.8 | 4.9 | 4.4 | 4.7 | 4.9 | 4.4 | 4.7 |
| | | | | | | | *Average* | 5.6 | 4.2 | 4.2 | 4.6 | 3.6 | 4.1 | 4.7 | 3.6 | 4.1 |

**Table 6**

**Empirical power that is obtained (in %) for the eight global homogeneity tests that are indicated (nominal error $\alpha$=5%)**

| $P_A$ | $P_B$ | $N_A$ | $N_B$ | $\pi$ | $O^+$ | $O^-$ | n | \multicolumn{8}{c}{Inference Method} | | | | | | | |
|---|---|---|---|---|---|---|---|---|---|---|---|---|---|---|---|
| | | | | | | | | d(a) | d(p) | LR | LR(a) | LR(p) | R | R(a) | R(p) |
| .7 | .8 | .7 | .7 | .35 | 5 | 2 | 100 | 5.5 | 4.6 | 7.8 | 3.8 | 3.6 | 8.1 | 4.0 | 4.0 |
| | | | | | | | 200 | 13.0 | 11.8 | 14.2 | 11.3 | 10.7 | 14.4 | 11.4 | 11.0 |
| | | | | | | | 300 | 19.7 | 18.7 | 20.8 | 18.3 | 17.9 | 20.9 | 18.5 | 18.1 |
| | | | | | 2 | 5 | 100 | 5.3 | 4.7 | 7.6 | 3.7 | 4.0 | 7.9 | 3.8 | 4.3 |
| | | | | | | | 200 | 13.4 | 12.5 | 14.7 | 11.5 | 11.5 | 14.9 | 11.6 | 11.8 |
| | | | | | | | 300 | 20.6 | 19.9 | 21.8 | 19.0 | 19.2 | 21.9 | 19.2 | 19.5 |
| | | | | .65 | 5 | 2 | 100 | 83.6 | 83.0 | 86.5 | 83.2 | 82.0 | 86.6 | 83.2 | 82.1 |
| | | | | | | | 200 | 99.7 | 99.7 | 99.7 | 99.7 | 99.7 | 99.7 | 99.7 | 99.7 |
| | | | | | | | 300 | 100.0 | 100.0 | 100.0 | 100.0 | 100.0 | 100.0 | 100.0 | 100.0 |
| | | | | | 2 | 5 | 100 | 88.4 | 88.1 | 90.8 | 88.1 | 87.4 | 90.8 | 88.1 | 87.4 |
| | | | | | | | 200 | 99.9 | 99.9 | 99.9 | 99.9 | 99.9 | 99.9 | 99.9 | 99.9 |
| | | | | | | | 300 | 100.0 | 100.0 | 100.0 | 100.0 | 100.0 | 100.0 | 100.0 | 100.0 |
| .7 | .8 | .8 | .8 | .35 | 5 | 2 | 100 | 15.3 | 14.6 | 16.6 | 13.1 | 13.8 | 16.7 | 13.3 | 14.1 |
| | | | | | | | 200 | 30.7 | 30.0 | 31.3 | 29.1 | 29.3 | 31.4 | 29.2 | 29.4 |
| | | | | | | | 300 | 45.1 | 44.5 | 45.5 | 43.8 | 44.1 | 45.6 | 43.9 | 44.2 |
| | | | | | 2 | 5 | 100 | 16.7 | 16.1 | 18.0 | 14.2 | 15.5 | 18.2 | 14.4 | 15.7 |
| | | | | | | | 200 | 34.5 | 34.0 | 35.2 | 32.6 | 33.4 | 35.3 | 32.8 | 33.5 |
| | | | | | | | 300 | 50.7 | 50.3 | 51.1 | 49.3 | 49.9 | 51.2 | 49.4 | 50.0 |
| | | | | .65 | 5 | 2 | 100 | 81.3 | 81.8 | 84.6 | 80.0 | 81.2 | 84.6 | 80.0 | 81.2 |
| | | | | | | | 200 | 99.3 | 99.3 | 99.4 | 99.3 | 99.3 | 99.4 | 99.3 | 99.3 |
| | | | | | | | 300 | 100.0 | 100.0 | 100.0 | 100.0 | 100.0 | 100.0 | 100.0 | 100.0 |
| | | | | | 2 | 5 | 100 | 87.6 | 88.3 | 90.2 | 86.6 | 87.8 | 90.2 | 86.6 | 87.9 |
| | | | | | | | 200 | 99.8 | 99.8 | 99.8 | 99.8 | 99.8 | 99.8 | 99.8 | 99.8 |
| | | | | | | | 300 | 100.0 | 100.0 | 100.0 | 100.0 | 100.0 | 100.0 | 100.0 | 100.0 |
| .7 | .8 | .7 | .8 | .35 | 5 | 2 | 100 | 78.5 | 79.9 | 80.8 | 76.3 | 79.5 | 80.9 | 76.4 | 79.5 |
| | | | | | | | 200 | 99.0 | 99.1 | 99.1 | 98.9 | 99.1 | 99.1 | 98.9 | 99.1 |
| | | | | | | | 300 | 100.0 | 100.0 | 100.0 | 100.0 | 100.0 | 100.0 | 100.0 | 100.0 |
| | | | | | 2 | 5 | 100 | 69.6 | 70.6 | 71.8 | 66.9 | 70.1 | 71.9 | 67.0 | 70.2 |
| | | | | | | | 200 | 96.9 | 97.1 | 97.1 | 96.7 | 97.0 | 97.1 | 96.7 | 97.0 |
| | | | | | | | 300 | 99.8 | 99.8 | 99.8 | 99.8 | 99.8 | 99.8 | 99.8 | 99.8 |
| | | | | .65 | 5 | 2 | 100 | 69.6 | 70.6 | 71.8 | 66.9 | 70.1 | 71.9 | 67.0 | 70.2 |
| | | | | | | | 200 | 96.9 | 97.1 | 97.1 | 96.7 | 97.0 | 97.1 | 96.7 | 97.0 |
| | | | | | | | 300 | 99.8 | 99.8 | 99.8 | 99.8 | 99.8 | 99.8 | 99.8 | 99.8 |
| | | | | | 2 | 5 | 100 | 78.5 | 79.9 | 80.9 | 76.3 | 79.5 | 80.9 | 76.4 | 79.5 |
| | | | | | | | 200 | 99.0 | 99.1 | 99.1 | 98.9 | 99.1 | 99.1 | 98.9 | 99.1 |
| | | | | | | | 300 | 100.0 | 100.0 | 100.0 | 100.0 | 100.0 | 100.0 | 100.0 | 100.0 |
| .7 | .8 | .8 | .7 | .35 | 5 | 2 | 100 | 96.0 | 95.6 | 97.4 | 96.2 | 95.0 | 97.4 | 96.3 | 95.1 |
| | | | | | | | 200 | 100.0 | 100.0 | 100.0 | 100.0 | 100.0 | 100.0 | 100.0 | 100.0 |
| | | | | | | | 300 | 100.0 | 100.0 | 100.0 | 100.0 | 100.0 | 100.0 | 100.0 | 100.0 |
| | | | | | 2 | 5 | 100 | 94.3 | 93.9 | 96.1 | 94.6 | 93.2 | 96.2 | 94.6 | 93.3 |
| | | | | | | | 200 | 100.0 | 100.0 | 100.0 | 100.0 | 100.0 | 100.0 | 100.0 | 100.0 |
| | | | | | | | 300 | 100.0 | 100.0 | 100.0 | 100.0 | 100.0 | 100.0 | 100.0 | 100.0 |
| | | | | .65 | 5 | 2 | 100 | 94.2 | 93.9 | 96.1 | 94.6 | 93.2 | 96.2 | 94.6 | 93.3 |
| | | | | | | | 200 | 100.0 | 100.0 | 100.0 | 100.0 | 100.0 | 100.0 | 100.0 | 100.0 |
| | | | | | | | 300 | 100.0 | 100.0 | 100.0 | 100.0 | 100.0 | 100.0 | 100.0 | 100.0 |
| | | | | | 2 | 5 | 100 | 95.9 | 95.6 | 97.4 | 96.2 | 95.0 | 97.4 | 96.2 | 95.0 |
| | | | | | | | 200 | 100.0 | 100.0 | 100.0 | 100.0 | 100.0 | 100.0 | 100.0 | 100.0 |
| | | | | | | | 300 | 100.0 | 100.0 | 100.0 | 100.0 | 100.0 | 100.0 | 100.0 | 100.0 |
| | | | | | | | *Minimum* | 5.3 | 4.6 | 7.6 | 3.7 | 3.6 | 7.9 | 3.8 | 4.0 |
| | | | | | | | *Maximum* | 100.0 | 100.0 | 100.0 | 100.0 | 100.0 | 100.0 | 100.0 | 100.0 |
| | | | | | | | *Average* | 76.6 | 76.5 | 77.5 | 75.9 | 76.2 | 77.5 | 76.0 | 76.3 |

# Table 7

## Some inferences based on the data from Table 1(a)

| | | 95%-CI | | | |
|---|---|---|---|---|---|
| | | Parameter | | | |
| Inference Method | Type | $d=P_A-P_B$ | | $\bar{d} = \bar{P}_A - \bar{P}_B$ | |
| d | Classic (liberal) | −0.0153 | +0.0410 | +0.0819 | +0.1922 |
| d(a) | Optimum | −0.0152 | +0.0411 | +0.0808 | +0.1907 |
| Inference Method | Type | Parameter | | | |
| | | $R=P_A/P_B$ | | $\bar{R} = \bar{P}_A / \bar{P}_B$ | |
| LR | Classic (liberal) | .9829 | 1.0473 | 1.1190 | 1.3116 |
| R(a) | Optimum | .9829 | 1.0475 | 1.1177 | 1.3096 |
| LR(a) | Almost equal to the optimum one | .9829 | 1.0475 | 1.1177 | 1.3096 |
| *Individual homogeneity test: statistic $z^2_{exp}$* | | | | | |
| | | Null hypothesis | | | |
| Inference Method | Type | $H_d$: $P_A=P_B$ | | $H_{\bar{d}}$: $\bar{P}_A = \bar{P}_B$ | |
| d | Classic (liberal) | 0.802 | | 23.73 | |
| LR | Classic (liberal) | 0.800 | | 22.44 | |
| d(p) | Optimum | 0.807 | | 22.50 | |
| R(p) | A little worse than the optimum one | 0.808 | | 22.32 | |
| d(a) | Optimum one for the compatibility test/*CI* | 0.809 | | 23.44 | |
| *Global homogeneity test: statistic $\chi^2_{exp}$* | | | | | |
| | | Null Hypothesis | | | |
| Inference Method | Type | $H_{d\bar{d}}$: $(P_A = P_B) \cap (\bar{P}_A = \bar{P}_B)$ | | | |
| d | Classic (liberal) | 25.94 | | | |
| R | Optimum | 24.45 | | | |
| LR | A little worse than the optimum one | 24.37 | | | |